# Exploring the added value of blockchain technology for the healthcare domain

*Lessons learned from two cases by a university hospital*

**November, 2019**

**By B.R.J. Bolmer, M. Taverne, M. Scherer**

# INTRODUCTION

Over the last few years the interest in blockchain technology has seen a significant increase. This was mainly due to the surge of the bitcoin rate, but also the development of alternatives to this coin and how the technology demonstrated its use for solving societal and organisational challenges (Zwitter & Herman, 2018). This left many organizations wondering if the technology could also be of added value for their processes.

In this article we would like to share our lessons learned from the past two years in which we, as a large university hospital employing over 13,000 employees, considered whether blockchain technology could be meaningful for healthcare and medical scientific research. The momentum created by the interest in the technology and the solution it presented, allowed for an exploration of its application in the healthcare domain (Agbo et al., 2019).

This article will expand on this exploration and will elaborate on how we tried to determine if the technology could help improve processes within the hospital. Through an open innovation program, two challenges were defined covering both patient care and scientific research. As part of the program, multidisciplinary teams were invited to try and find solutions which were later tested in a pilot at the hospital. Before further elaborating on how this was organized, we will briefly touch upon why blockchain technology should be considered (without becoming too technical), before moving on to the challenges that were set out to validate it.

Ever since the introduction of the bitcoin, blockchain as a technology has improved significantly. Back in 2010, when the coin was on the market for a year, there were already some wild ideas of buying products and services with the coin as a currency. For example, on May 22$^{nd}$ that year, someone bought two large pizzas for approximately 10,000 bitcoins, worth 30 dollar at that time. If you would buy those two pizzas today, you would be paying approximately 1/300$^{th}$ of a bitcoin.

While the rate of the bitcoin has been extremely volatile, the increase of its use, and therefore validation of the underlying technology, has led to improvements we can see today (Miraz & Maaruf, 2018). A lot of alternatives to bitcoin have been developed, of which Ethereum has become the most well-known. These alternatives, each having their own (unique) set of characteristics, demonstrated that it was more than a currency. Logic has been added and variations showed how blockchain technology could be used for different purposes, while staying true to what it was originally developed for (Sherman et al., 2018).

By design, blockchain technology provides trust, transparency and immutability on transactions in a network (Sultan et al., 2018). It can be compared to a ledger that records all transactions ever performed by two or more parties. The ledger is open and distributed among everyone that participates in the network, providing a single source of truth without the possibility to revert performed transactions. This is guaranteed using cryptography; all recorded transactions are dependent on each other and making alterations would also affect all subsequent transactions. Based on an agreed protocol, new transactions can be validated by all participants and can be added to the chain of blocks holding other submitted transactions (Vujicic et al., 2018). This validation removes the need for a trusted
.

third party to manage each transaction, making the network act independent (e.g. for transferring bitcoins between individuals, no intermediate bank is necessary to validate the transaction).

For healthcare providers, such as the university hospital, these characteristics of trust, transparency and immutability are also important from an organizational perspective. This becomes increasingly apparent when looking at the types of sensitive data that are being stored (Dubovitskaya et al., 2017). A current debate on this perspective is one of data ownership; which raises the question whether patients should become more in control over their data, how they can get insights on how their data is being used and how they can manage data permissions (Gordon & Catalini, 2018). The way blockchain technology implements the aforementioned characteristics can provide solutions for these and other challenges that might be difficult to solve otherwise.

To explore the possibilities of blockchain technology, the university hospital participated in an open innovation program organized by Odyssey (Odyssey, 2019). As part of the innovation program, the hospital provided two challenges for which blockchain may be a viable solution; one for patient care and one for scientific research. The goal of these challenges was to determine if the technology as an innovation could provide a fitting solution. A blockchain innovation workgroup was established by the hospital to oversee the coordination of these two challenges and the participation of the hospital in the program. The workgroup attended sessions that looked at legal, ethical, and technical aspects of blockchain, organized a session on applying blockchain within the healthcare domain, participated in the blockchain hackathon, collaborated with the participating hackathon teams to develop a prototype, and performed a pilot project to validate the solution.

In the following chapters of this article, we will first consider literature that has looked into the application of blockchain technology within the healthcare domain. Then, we will get into more details on the participation in the innovation program and how we came to two cases to validate the technology. Thereafter, we will describe the results from these cases, the intended pilot and the impact of the innovation as it brought change to the organizations involved. We will conclude with a discussion; looking back at the results of the challenges and the future of innovation within the hospital.

## BACKGROUND AND SIGNIFICANCE

### Established architecture and way of work

It is becoming increasingly important to perform diligent data management of healthcare data, which includes storage, access control, and sharing of data (Mulder, 2019). With the expansion of possibilities to use and interpret data, it is no longer only limited to creating information for healthcare practitioners who use the data to determine the best possible treatment for patients. A shift is noticeable in which this traditional relationship is changing and the patient is empowered to do his or her own research (Panda et al., 2019). To facilitate this, healthcare providers need to rethink how they are managing sensitive patient data and need to consider if their current IT-landscape can enable this. When done

.

correctly, proper management of healthcare data "improves healthcare outcomes by allowing holistic views of patients, personalized treatments, and efficient communication" (Katuwal et al., 2018).

By recently implementing an Electronic Patient Records system (EPRs), the university hospital has taken a big step in facilitating patient empowerment. The next step is to better support self-managed healthcare and provide a Personal Health Records system (PHRs) for anyone to access. To develop such a platform for patients, but also healthy individuals, a lot of other organizations have to join-in in order to create a comprehensive view of the data (Azarm-Daigle et al., 2015). This includes, but is not limited to, general practitioners, pharmacists, specialists and other hospitals, but also regulators, governments and insurance companies. All of these organizations store data and information on an individual that contributes to providing healthcare. For most people it is often unclear which organizations are storing their data, much less that they know for what reason it is being used (Wetzel et al., 2018, Caine & Hanania, 2013). This could lead to incomplete records, possibly making them inaccurate, which could lead to unbeneficial medical decisions.

Besides contributing to patient care, making data more widely accessible and doing so transparently can also benefit scientific research. Stepping into a hospital and undergoing medical treatment will also subject the data from the treatment to scientific research done by the practitioners and researchers involved. With regulations such as the General Data Protection Regulation (GDPR), healthcare providers are mandated to inform patients on how their data is being used. By incorporating this into a platform such as a PHR, studies that are not directly related to the performed medical treatment will also have a better possibility to recruit their study population (Wilcox et al., 2009, Hackett et al., 2019). Ideally, everyone that would like to participate in scientific research, should be able to decide how they would like to make their data available.

**Current limitations and reasons for change**

Ekblaw et al. (2016) emphasize a need for change by stating that "a long-standing focus on compliance has traditionally constrained development of fundamental design changes for [PHRs]. We now face a critical need for such innovation, as personalization and data science prompt patients to engage in the details of their healthcare and restore agency over their medical data." Over time, individuals leave their data "scattered across various organizations as life events take them away from one provider's data silo and into another." In doing so it is difficult to access personal historical data, as it is often the healthcare provider who retains primary stewardship.

The collaboration required between different providers and their information systems create interoperability challenges that may "pose additional barriers to effective data sharing" (Gordon & Catalini, 2018). Without coordinated data management and exchange the records remain fragmented, rather than cohesive. The result is that records staying siloed within the source system, which is seldom designed to effectively share data. This often leads to the creation of alternative solutions, in which organizations that would like to share data each develop their own platform, burdening the individual with requests for permissions to share data via each platform. For most organizations "an individual patient's
.

health data is [still] scattered across numerous systems, and no institution has a complete picture" (Gordon & Catalini, 2018).

The university hospital is aware of the current challenges and states that the "exchange of information between various healthcare providers, between the healthcare provider and the patient, the introduction of PHRs, the supply of data for quality registers, the use of data for process optimization and research, and the use of data for precision medicine each have their own set of requirements" (UMCG, 2018). Such increasing complexity in the use of data opens up the opportunity to explore solutions that would otherwise, in a more stable context, remain unconsidered. Given the fundamentals of blockchain technology, its application could be a solution to some of the current limitations (Agbo et al., 2019).

**Why use blockchain technology?**

Data being used for medical care and research, such as records stored in EPRs and PHRs, are "critical, highly sensitive private information, and need to be frequently shared among peers" (Dubovitskaya et al., 2018). More so, the sharing of such data may also extent to other organizations, outside of the control of the original source. Developing a system that is designed to share such sensitive data can be a challenge, as organizations are often (rightfully so) very protective over their data (Azarm-Daigle et al., 2015).

Using blockchain technology as part of a solution can prove to be a good way to address these concerns, as "trust and traceability are the two basic promises of the blockchain obtained out of the box which solves the generic trust problem on all public, federated, and organization levels" (Katuwal et al., 2018). Trust is accomplished because the technology "provides a shared, immutable and transparent history of all the transactions" (Dubovitskaya et al., 2018). The resulting log, containing all of the transactions performed by all of the involved stakeholders, makes it possible to enforce accountability for each stakeholder individually.

This alone, however, is not always sufficient to "provide a complete solution, which is why we often see blockchain paired with strong cryptographic protocols" (Katuwal et al., 2018). One example is the use of zero-knowledge proofs, which is applied to prove correctness and validity of a transaction, without revealing the actual content. In other words, you share information about a secret, but not the secret itself (Grzonkowski & Corcoran, 2014). Security is provided as "data recorded in the blockchain cannot be changed or deleted without leaving a trace. This immutability and traceability of the data is a critical requirement for any health care system" (Katuwal et al., 2018)

On a higher level, "blockchain technology can be thought of as a platform for digital exchange, where the platform functions without a traditional intermediary. Health data can live in multiple systems and sharing data requires numerous points of collaboration between entities. As interoperability becomes more patient-centric, there is an opportunity to leverage blockchain technology to facilitate this exchange and give patients greater control over their data." (Gordon & Catalini, 2018). Such developments will provide patients with the benefit of a comprehensive, transparent picture of their medical history, which is

.

important in "establishing trust and continued participation in the medical system, as patients that doubt the confidentiality of their records may abstain from full, honest disclosures or even avoid treatment" (Ekblaw et al., 2016).

Important to note is that solutions that apply blockchain technology have to consider regulations on how to process sensitive data, such as an individual's right to be forgotten. One hotly debated topic is whether hashed and encrypted data can still be seen as personal data. In a recent report by the European Parliament (2019) it is stated that "whereas it is often assumed that this is not the case, such data likely does qualify as personal data for GDPR purposes." It is therefore not desirable to store actual health data on the blockchain. Instead, solutions should facilitate the data exchange by providing an audit log, managing authorizations, or referencing to other (off-chain) systems. Even then, it is imperative that solutions using blockchain technology have to be safe, reliable and dependable. This should not only support in the adoption by patients, but also healthcare practitioners, and therefore increase its usefulness (Brogan et al., 2018).

## LEARNING BY DOING

As the university hospital we set out to explore the added value of blockchain technology for the healthcare domain. Since this was no core business of the hospital, and prior knowledge on the technology was limited, we partnered with an innovation program that provided a hackathon as a means to kick-start projects. In preparation of this hackathon we formulated two compelling challenges for participating teams to work on. The participation of the hospital in the program was organized by the established blockchain innovation workgroup.

Starting with a vision on how to leverage the specific traits of blockchain, we back-casted to what we required today. In defining the use-cases we came across six hospital processes in need of a solution in line with our vision. Based on input from experts and those involved, we selected the most suitable use-cases. Besides looking at the fit between the use-case and the technology, we also considered favourable conditions to, for example, perform a pilot and demonstrate developed applications and solutions.

In anticipation of the hackathon we prepared a toolkit including all relevant information on the use-cases and challenges, made sure necessary (test) data was available, as well as stakeholders, experts and end-users for each challenge. A few weeks before the hackathon, problem-owners presented their case to the participating teams to give them an opportunity to ask questions. This gave us the chance to verify if the toolkit was ready to be used. Additionally, we prepared coaching and judging sessions during and after the hackathon (which was dubbed the acceleration program for promising teams) to support the teams.

During the three-day hackathon ten teams worked on the two challenges. The blockchain innovation workgroup made sure they had access to the right people at the right time. For example, business developers only came in on the second day in the afternoon, but were also available for call if teams had any questions. We have also involved the end-users (such as patients) and invited them to attend the hackathon to help the teams. During the

.

weekend we did rounds to learn about the developments of all of the teams and to pre-judge their solutions, provide feedback and answer questions. This assured that we were not depending on a snapshot final presentation for evaluating the potential of the solutions and the teams.

A month after the hackathon we organized the acceleration meetup and invited four promising teams to present their solutions to stakeholders and investors. This event provided an opportunity for the teams to explore future collaborations and even determine if solutions could be complementary. Based on a mix of (favourable) conditions within the hospital, the feasibility of the solution and team cohesion, two teams were selected to each work on a project to demonstrate their solution. The cooperation with the teams was secured by a contract and allocating resources from the hospital to the co-creation of the solutions designed by the teams. Unfortunately, due to unforeseen personal circumstances, one of the teams had to back out of the project and only delivered the project plan. The other team continued to develop their solution in an agile manner with updates and feedback moments every other week.

After about six months of development we partnered with the dHealth lab. Via this lab we were able to test the application in a scripted simulation with students, before demonstrating the final solution to professionals and patients. This simulation supported the adoption of the application as we wanted to prevent any technical and logistical problems from occurring, before going into production and using the application with actual patients. Unfortunately, unforeseen circumstances threw a spanner in the works as the program and the use-case for which we wanted to introduce the application was no longer available for piloting. Nevertheless, the way in which the initial challenge was set up, and how the application architecture was developed, gave us the opportunity to revisit five other relevant use-cases for which the developed application could also prove to be useful.

In conclusion, exploring the added value of blockchain technology for the healthcare domain has not only taught us about the technology, but also about the unpredictable course of innovating, and how to use an innovation program to successfully pan out ideas and come to a product. Our main lessons learned from this process are:

- Use a hackathon [insert any other form of pressure cooker] as a (design) sprint within a complete innovation process.
- Set a challenge based on a vision, cast back to what needs to be done today, find a committed sponsor, suitable use-case and check for favourable conditions in order to start iterative development after the pressure cooker.
- Find sponsors who are involved in the primary process that is being affected by the change and/or innovation, or who support the main cause. Having commitment from those who are affected, and where added value is provided, is important.
- Make sure end-users, stakeholders, experts, systems and data are present/accessible before, during and after the pressure cooker. Allow teams to ask questions and provide feedback on the challenge and the input you present.

.

- Make sure there are multiple moments to check the progress of the teams during a pressure cooker in order to get a good idea of how well a team is functioning, check execution power and not let your judgement depend on one single presentation.
- Have everything prepared for an acceleration program after the pressure cooker, include all stakeholders, invite all teams and solutions that show potential and steer for decision making and practical follow-up actions.
- Have regular feedback sessions with all stakeholders involved.
- Simulate the final solution during the development in order to find out about all conditions that need to be present in the real life (pilot) setting for it to be successful.
- Use public events to present progress; providing (hard) deadlines helps the project move forward. This mitigates scope creep by selecting the features that are really necessary, as the possibilities for development are often endless and results are never perfect.
- When encountering unforeseen circumstances, be transparent about it to all parties involved as soon as possible. Explore alternatives together within a predetermined period of time and decide whether or not to proceed. Evaluate the first initiative and bring it to a closing, before continuing fresh with all the lessons learned.

## CASES AND RESULT

### Patient care at home

*Case description*

For our patient-centric case, we selected a program that helps vulnerable elderly patients in their transfer back home, the day after being admitted to the hospital due to an emergency situation. From their own home, they receive care from the hospital in cooperation with other care providers. Research has shown that recovery of these patients improves when they are being treated in a comfortable, familiar environment (Pouw et al., 2018).

To be able to provide qualitative good care at the patient's home, it is important that all involved healthcare practitioners have the most up to date data on the patient's well-being. They should be able to get information on the relevant treatment, have the possibility to add treatments and conditions, and receive updates related to the treatment from, for example, other practitioners. This information should be presented via a viewer that provides a comprehensive and concise overview.

In the current situation this is accomplished by a paper record at the patient's home, in which anyone who provides care can record their actions, observations, and concerns. One of the disadvantages of this method is that these recordings also have to be registered into the EHR-systems from each organization, creating extra administrative burden for the healthcare practitioners. Additionally, if a practitioner would like to obtain the latest information on the patient, they are required to go to the patient's home to look into the records. This could create a situation where information is not delivered on time to the right professional. Finally, privacy-wise, it is not desirable that everyone who visits the patient, authorized or not, is able to view the information of the treatment.

.

Up till now we have not had the opportunity to enable electronic data sharing between the organizations in an effective and efficient way. Blockchain technology might prove to be useful for this case, considering its characteristics, and may pave the way for implementation. To determine if this is the case, the following question has been proposed:

a. *"Help us create a safe, private, efficient and up to date health record to deliver (hospital) healthcare at the patient's home with multiple healthcare institutions."*

For the solution to be viable, several aspects of the data sharing process had to be considered. For example, data sharing has to happen real time and the risk of downtime and losing data should be absolutely minimal. In case of an emergency, relevant data should (directly) be accessible for the practitioner supporting the patient and should not prevent treatments. Because the healthcare program is still developing and also expanding, scalability is important as other healthcare providers need to be added on request.

*Proposed solution and implementation*

To develop the solution, we determined a set of functional and technical requirements that had to be demonstrated by the application. One of the main requirements was that the application needed to support real-time sharing of data between various healthcare providers and different EHR-systems. Since time is an important variable (in, for example, medication intake or treatment), this data had to be visualized in a logical timeline. In order to view this data, the application had to provide the advantage of being available at any location, as opposed to the current paper record.

Another requirement was the scalability of the application and the ability to easily add other organizations to the healthcare program. For this aspect particularly, we were interested in how blockchain technology could prove to be useful. One of the advantages of the technology is that it is distributed. In order to achieve this, sufficient organizations (or individuals) have to be involved in the network. Looking at public blockchain networks, each individual who joins the network also becomes part of the distribution. This would, however, not be suitable for this application, as sensitive data could not be distributed to unauthorized individuals. Instead, we employed permissioned blockchain technology to maintain and governance access control policies. The permissioning creates more control over who can join the network, as opposed to a public network which is open to anyone.

In order to add organizations to the network, each new organization receives its own blockchain node; a server containing the ledger and an application programming interface (API) specified for the organization to connect with its own healthcare system. Having such a node would allow for a governance model in which data requests can be performed from within the perimeter of the organization. To perform a transaction and retrieve data, two variables are required. The first is the requester, which can be the healthcare practitioners from the hospital submitting a request to, for example, view data which is controlled by the home care organization. Second is the sender, which is the organization who receives the request and needs to do the API-call on its own system. In this example, this would be the home care organization. After authorizing the request, the retrieved data is sent to the
.

application for the practitioner to see. In figure 1 these transactions are outlined in a sequence diagram.

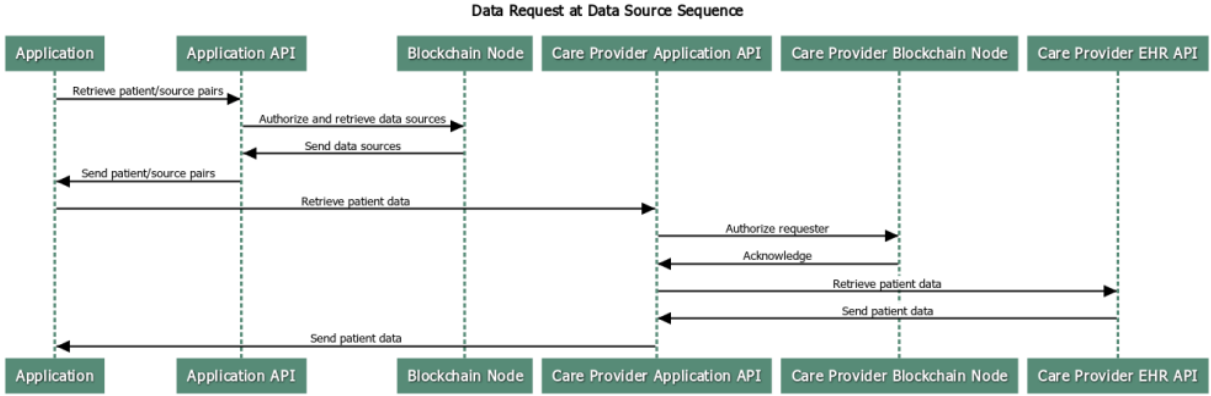

Figure 1: Sequence diagram showing the transactions performed by the application, blockchain nodes and EHR-systems from a care provider (i.e. data source) (made by Lemon Care)

Using a (smart) contract that is also published on the network, healthcare practitioners are linked to their own organization and a corresponding treatment plan that belongs to a patient. Using this as a basis, and knowing which organizations are part of the treatment plan, the patient can authorize his or her caregiver to access the relevant healthcare records. Based on this authorization, healthcare practitioners will be able to successfully perform a request at a different organization. This approach also ensures that the data does not have to be transferred to a shared database for access, but that it can be presented in the application for a specified duration (e.g. the time a practitioner requires to update the treatment). Subsequently, by recording these requests in transactions on the network, patients will be able to gain insights into who accessed their data and at which time.

For the first version of the solution we focussed on the healthcare practitioners and verified with these users how they wanted to have the data visualized. Some examples of this can be seen below.

.

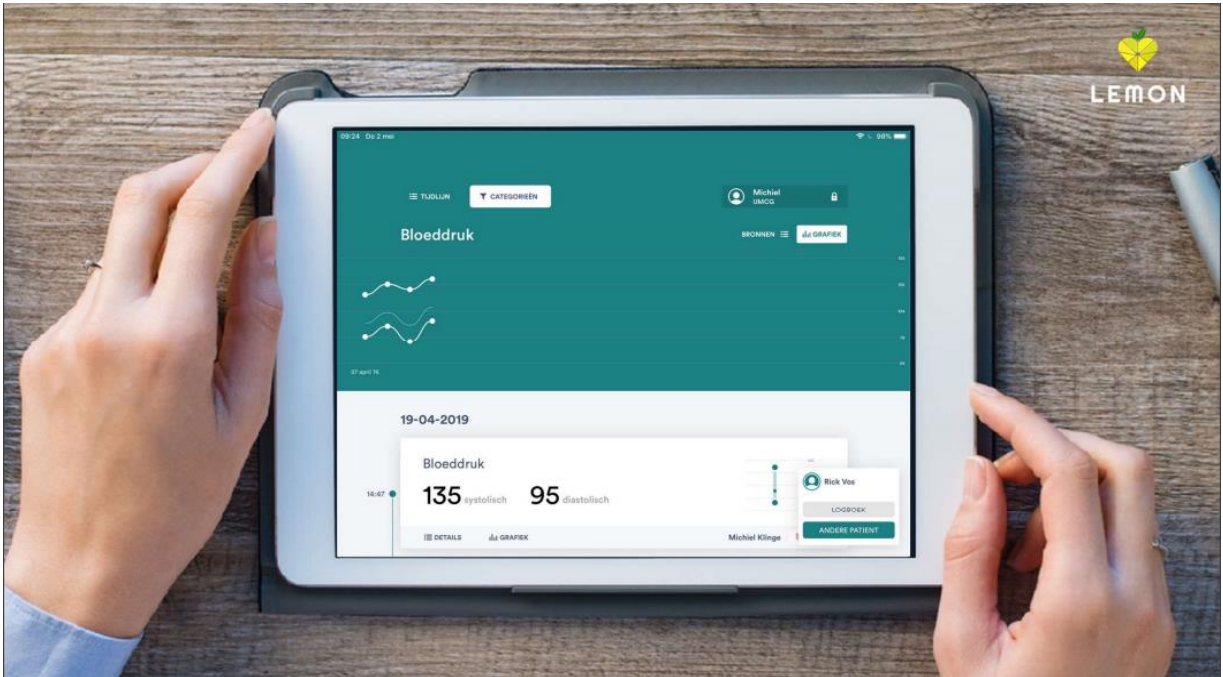

Figure 2: Visualizing trends in measurements (e.g. blood pressure) for patients and healthcare practitioners within the application (made by Lemon Care)

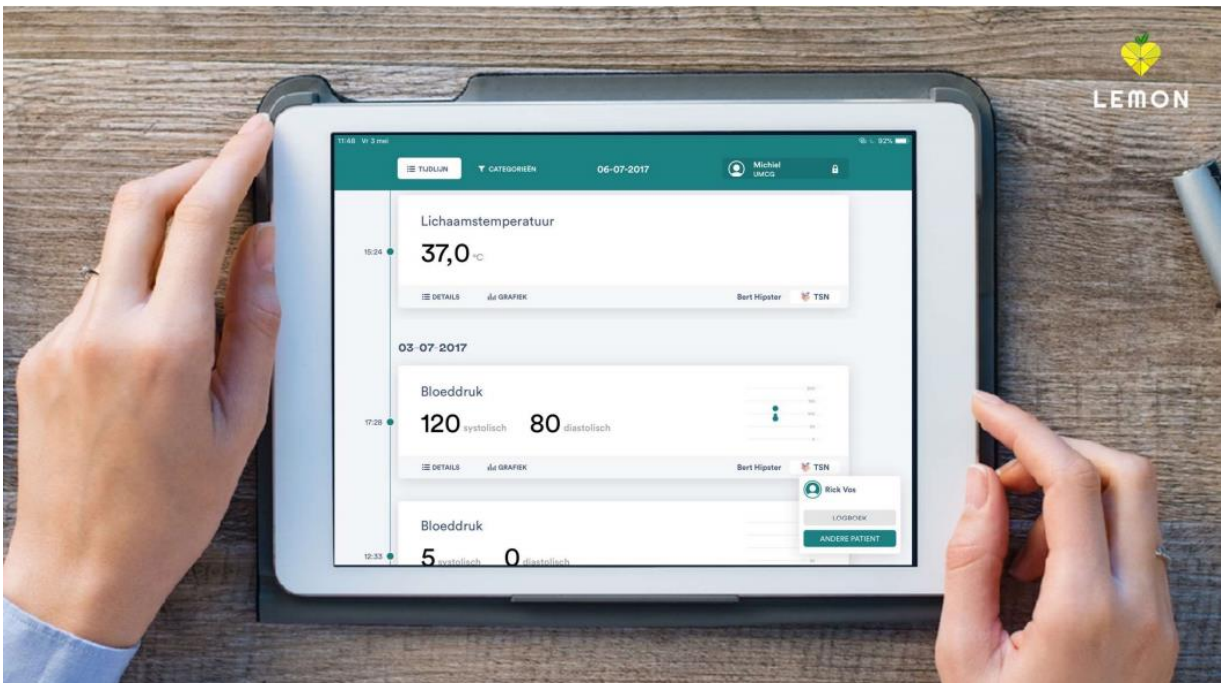

Figure 3: Multiple measurements from different sources available, sorted on time (made by Lemon Care)

.

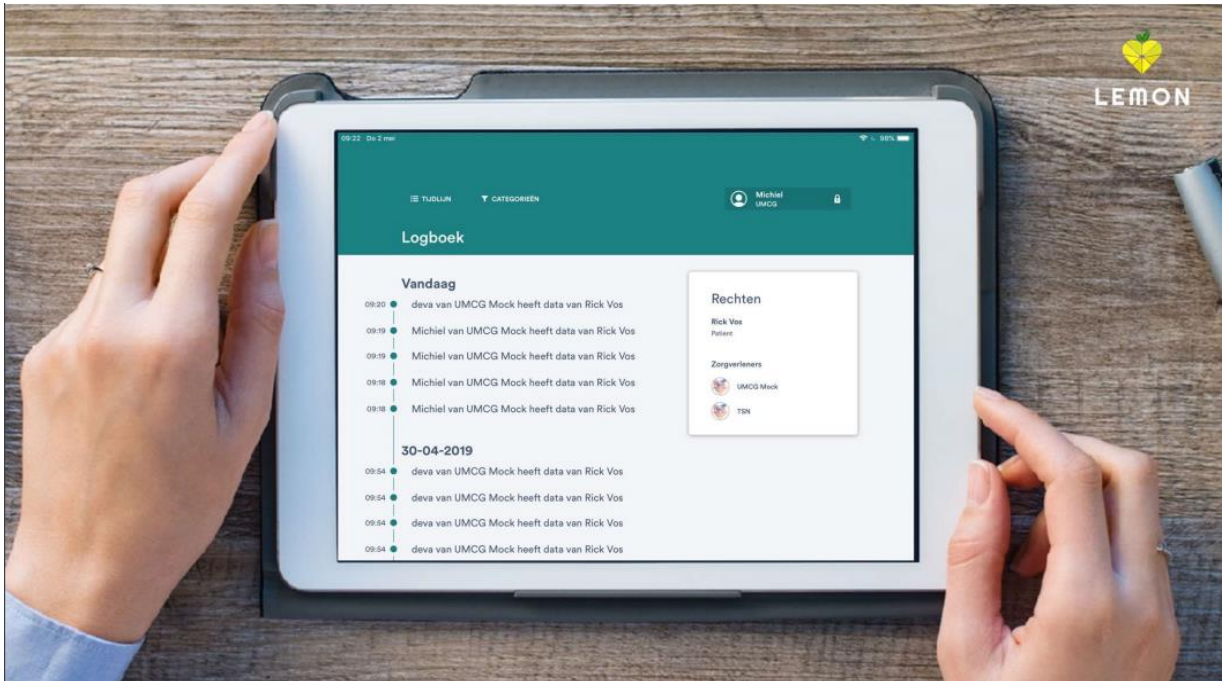

Figure 4: Log showing all of the performed transactions on the data (e.g. permission to share and access to data) (made by Lemon Care)

**Consent management for scientific research**

*Case description*

Scientific research advances the way we can perform treatments and is an important part in laying the fundamentals for unraveling diseases, curing patients and learning how we can stay healthy. The input for doing such research is data; from patients but also from healthy individuals. Researchers can obtain data by either recruiting their own study population or by requesting it from, for example, a biobank. Via a biobank, data is collected based on protocols to guarantee sound data and create the possibility for longitudinal research. By using data from such a source, researchers are able to compare individuals in the same manner. While this results in a lot of useful data, it can still be difficult to collect data from a very specific research niche (such as a rare disease).

For this case we looked at individuals that experienced difficulties in sharing their data for research and healthcare. One of the relevant use-cases was from a patient and biobank participant, who was unable to share his medical data in such a way that he could reach out to the right hospitals and researchers. The patient suffered from a spinal condition resulting in back problems. According to the doctors helping him, all treatment options were exhausted. This made him decide to publish all of his medical data online, on a personal website, hoping that someone would be able to find a fitting treatment.

Nowadays we collect a lot of data on our own (quantifying ourselves), but also request more openness and transparency from the practitioners who treat us (by obtaining insight into our medical records). This raised the question on how we could assist individuals in making their data available for both finding treatment possibilities and supporting scientific research. The current possibilities to share data for scientific research are limited.
.

Therefore, we would like to determine if we can use blockchain technology to enable the sharing of data between individuals and researchers:

b. *"Help us with creating a safe and irrefutable way of sharing data with researchers, in which the individual has ownership of his data and has insight into how his data is being used"*

The solution aims to bring the researcher and the individual closer together. Before the data can be used for scientific research, the individual will have to provide his or her consent for the aim of the research. In order to increase and guarantee participation, researchers require a novel way to inform individuals about the research and ask for their consent, which should then also be legally binding. In addition, individuals and researchers need to be linked to each other, as it is important that the right data is provided to the corresponding study. In turn, participants might become more engaged if they can keep track of the results of the study and learn about the conclusions drawn from the data, which could contribute to the patient's willingness to participate in (other) studies.

*Proposed solution and implementation*

After learning from the case and discussing the need for a solution with the patient, we sat down with researchers to further define the requirements. The focus for the first implementation of the solution was the consent collection, which is necessary before individuals can participate in research and share their data in a useful way. In order to improve the consent management process, various challenges had to be addressed. The application had to support individuals in understanding the consent they are giving. It also had to ensure that the consent was provided to the researchers on time and is matched to the corresponding participant.

As a solution, a dashboard for researchers was proposed in which a study could be created through which individuals interested in the study could give their consent electronically. The consent was presented to individuals as an informative quiz, in which several questions were asked to confirm that the individual knew what was being asked from him or her when participating in the study. The view that was shown to the participant also served as a means to reaffirm the study, show the researchers involved, and provide an explanation on the (possible) study outcomes.

Blockchain technology was used to create an audit trail of the process of consent collection, starting from the moment the consent invitation was sent to the participant. Additionally, if the participant was able to complete the consent procedure (i.e. the quiz), he or she could sign it digitally. This step would then also be recorded, which would give the researcher direct insight into the validity of the consent. Through the dashboard the researcher would also see if the participant struggled with any questions in the quiz, or how many mistakes were made (to prevent participants from - just - guessing the answers to sign the consent). Using this information, the researcher could verify if the participant understood the consent, and helped determine if additional explanation was required.

.

On a higher level, the application could serve as a match-making platform, allowing individuals to first create a meta-data profile containing the sources of their data (e.g. a biobank). Through a multi-layered consent mechanism, containing the consent for the meta-data profile and an actual study, an individual would gain more control over what he or she is sharing. This could be adjusted on the level for all researchers to see, but also per study. Researchers would then be able to query the meta-data profiles of individuals first (through f.e. a zero-knowledge proof protocol) to find their initial study population, before contacting possible participants individually for the actual consent.

For the first version of the consent collection for participants this was the design of the (quiz) viewer:

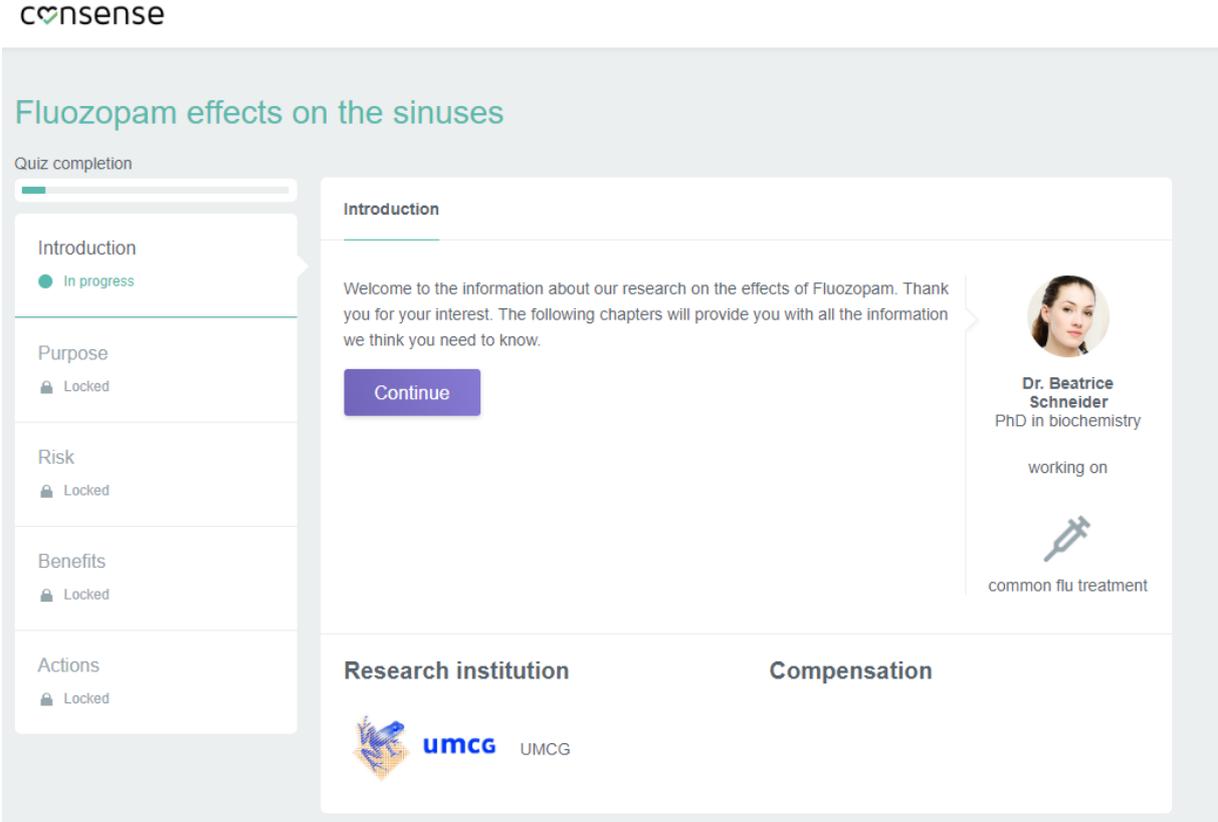

Figure 5: introduction to a research study for participants. On the left participants can go through the quiz created by the researcher (made by Consense Data Exchange)

## DISCUSSION AND CONCLUSION

By looking at two cases present in the university hospital we were able to learn more about how blockchain technology can contribute to solutions that benefit healthcare and medical scientific research. While the case of consent management remained conceptual, the project for patient care at home resulted in an application supported with blockchain technology. This application was successfully demonstrated through a simulation and had enough stakeholder support to continue implementation.

.

We have seen that some of the advantages of blockchain technology can reinforce key features of an application, such as having a distributed architecture to facilitate data sharing between multiple parties. Furthermore, by having more fine-grained information on who accessed which data at what time, authenticity of transactions can be verified and guaranteed. At the same time, the limitations of the technology should not be disregarded, especially when comparing blockchain to more traditional (database) technologies. For example, storing sensitive data on the blockchain or references to any personal information is undesirable, as data recorded on the network can not be altered or removed. The application of blockchain technology should therefore be considered in its context, as it is not a replacement for other database applications that could also prove useful.

For the immediate future in data management and patient empowerment, Gordon and Catalini (2018) state that "as data liquidity becomes less of a concern through expanded APIs, and as patients obtain better electronic access to their data, they can increasingly become the digital stewards of their health data. The data may still be largely generated in institutional silos, but patients will now have the ability to build a comprehensive view of their health, retrieving their data and sharing it as appropriate with other entities." Looking at the developments of blockchain over the past two years, we see a role for this technology in specific cases to leverage the added value.

Through the innovation program we were able to demonstrate that blockchain technology has the potential to contribute to solving problems in the healthcare domain. This supports earlier findings showing that the technology can help set-up a "patient-centric healthcare model", but can also be applied more practically by minimizing "vendor lock-in problems in healthcare" (Katuwal et al., 2018). Nevertheless, it is important to remain critical, as blockchain technology still faces challenges regarding privacy, regulations and standards and healthcare in general, such as data storage and data anonymization (Dubovitskaya et al., 2017). In addition, the scope of the discussed cases does not cover the entire domain (including f.e. laboratory activities or care expenses). To tackle these issues, we invite others to take a similar approach and learn from innovation to break new ground.

In conclusion, starting from a vision, defining two challenges supported by relevant use-cases, supporting teams dedicated to solving the problem at hand, involving relevant stakeholders and building a solution, has demonstrated that it was worthwhile to explore this technology. Considering that there was little to no knowledge about blockchain within the organization, and having a working prototype at the end of the two years, also shows that such a learning process can have actual benefits to an organization such as the university hospital.

Apart from having the new technology demonstrated, the exploration process also created momentum for change. Together with the universities in the area, the municipality and the province, the university hospital is continuing the innovation program under the umbrella of the patient-centric innovation department (PACE). The hackathon will serve as a catalysator to increase collaboration through a joint challenge on how to ensure affordable and accessible regional acute care.

.

# ACKNOWLEDGEMENTS

The authors would like to thank the following organizations for supporting the innovation workgroup and help explore this innovative technology and its application for the university hospital: Odyssey, TSN Thuiszorg, Martini Ziekenhuis, Hospital@Home, Lemon Care, Consense Data Exchange, dHealth lab, Ministry of Health, Welfare and Sport, Lifelines, Pels Rijcken, Zorginstituut Nederland / Mijn Zorg Log, CareChain, and Ledger Leopard. Furthermore, the authors would like to thank all individuals who contributed to this project and its results.

This project was performed by the University Medical Center Groningen (UMCG).

# ABOUT THE AUTHORS

B.R.J. Bolmer works for Lifelines as a technical data manager and has expertise in data management and a background in computer science. M. Taverne works as an innovation consultant for the function of Patient Centered Innovation & Connected Care within the UMCG, facilitating (digital) innovation projects and strengthening the innovative capacity of the hospital. M. Scherer is CTO of the UMCG and has responsibility over the department of Medical and Information Technology, coordinating all IT-related projects

.

.